\documentclass[%
preprint,
superscriptaddress,
 amsmath,amssymb,
 aps,
prb,
]{revtex4-1}

\usepackage{graphicx}% Include figure files
\usepackage{dcolumn}% Align table columns on decimal point
\usepackage{bm}% bold math

\usepackage{textcomp,gensymb} %added by GVK for creating symbols like degree
\usepackage{xcolor} %added by GVK 
\usepackage {graphicx} %added by GVK 
\usepackage [caption = false] {subfig}
\newcommand\newGVK[1]{{\color{black} #1}}

\begin{document}

%\preprint{APS/123-QED}

\title{Study of energetics of 360$^{\degree}$ domain walls  through annihilation}% Force line breaks with \\

\author{G.V. Karnad}
\affiliation{Institut f{\"u}r Physik, Johannes Gutenberg-Universit{\"a}t, Mainz, Germany}

\author{E. Martinez}
\email{edumartinez@usal.es}
\affiliation
{Departamento Fisica Applicada, Universidad de Salamanca, Salamanca,
Spain}
\author{M. Voto}
\affiliation
{Departamento Fisica Applicada, Universidad de Salamanca, Salamanca,
Spain}

\author{T. Schulz}
\affiliation{Institut f{\"u}r Physik, Johannes Gutenberg-Universit{\"a}t, Mainz, Germany}

\author{B. Ocker }
\affiliation{Singulus Technology AG, Kahl am Main, Germany}

\author{D. Ravelosona}
\affiliation{Centre for Nanoscience and Nanotechnology, University Paris-Saclay, Orsay, France}

\author{M. Kl{\"a}ui}
\email{klaeui@uni-mainz.de}
\affiliation{Institut f{\"u}r Physik, Johannes Gutenberg-Universit{\"a}t, Mainz, Germany}

\affiliation{Graduate School of Excellence \textquotedblleft{Materials Science in Mainz}\textquotedblright (MAINZ), Mainz, Germany}

\date{\today}% It is always \today, today,

\begin{abstract}

The Dzyaloshinskii-Moriya interaction (DMI) causes domain walls in perpendicular magnetized systems to adopt a homochiral configuration by winding in the same direction for both Up-Down and Down-Up walls. The topology of these domain walls is then distinct from the uniformly magnetized state. When two domain walls approach each other and are in close proximity they form winding pairs, stabilized by a dipolar repulsion. This can result in the formation of 360 $^{\degree}$ stable domain walls, whose stability is directly related to the magnitude of the additional dipolar interaction resulting from the spin structure governed by the DMI. Application of an external magnetic field can overcome the dipolar repulsion of the winding pairs and result in the annihilation of the domain walls, which is studied here in a combined theoretical and experimental effort. We present an extended analytical model that studies the interaction and modification of the dipolar interaction of the domain wall pairs under the application of in-plane and out-of-plane magnetic fields. We realize the experiment in a system of  Ta$\mid$Co$_{20}$F$_{60}$B$_{20}$$\mid$MgO and observe that the results are in agreement with the behavior predicted by the analytical model. To compare and understand these results, we perform micromagnetic calculations to gauge the validity of the analytics and also include the full dipolar interactions which are present due to the device geometry. We find that our numerical and experimental studies are in agreement and that the DMI indeed provides an additional stability mechanism against annihilation of DWs, which is potentially useful in  dense memory storage applications. Beyond implications for domain walls, understanding the  interaction is an important step to understand and control the interaction of many spin structures that contain domain walls, such as skyrmions.

\end{abstract}

\maketitle

%\tableofcontents
\section{Introduction}
Spin-orbit interactions in magnetic multilayers have in recent years generated tremendous interest due to the vast possibilities \cite{parkin2015memory,fert2017magnetic} of optimizing and tailoring them for technological applications. These applications mainly rely on stable magnetic textures and their manipulation - either by magnetic fields or electric currents. One of the key ingredients to stabilize these spin textures is the interfacial Dzyaloshinskii-Moriya interaction (DMI) \cite{DZYALOSHINSKY1958241,Moriya,fert1990}. DMI has been predicted \cite{fert1990,thiaville2012dynamics} and observed \cite{bode2007chiral} in systems with broken inversion symmetry. One such example is that of a ferromagnet (FM) sandwiched between a heavy metal (HM) and an oxide (Ox) layer. This results in a system such as HM/FM/Ox, which possesses no structural inversion symmetry.

DMI is an antisymmetric exchange interaction and its energetics can be described as an additional term in the Hamiltonian \textbf{H$_{DMI} = - $D$_{ij}$ $\cdot$  (S$_{i}$ $\times$ S$_{j}$)}, where D$_{ij}$ is the DMI vector and S$_{i}$ and S$_{j}$ are two neighboring spin moments. The vectorial nature of the DMI induces a fixed handedness or winding to the transition between the Up-Down/Down-Up magnetization because it lifts the energetic degeneracy of domain walls with different handedness. This can be characterized by its topological charge or winding number \cite{Braun2012}, $w_n$ = $\frac{1}{2\pi}\int_{x_1}^{x_2}  \partial_x \alpha $ $ dx$ \newGVK{in a one-dimensional sense} ($\alpha$ is the angle between the spin and the film normal, $x$ indicates the coordinate axis through the wall, $x_1$ and $x_2$ are positions within the domains). Since the DW winds only halfway around the circle in spin space one finds a value of $\vert w_n \vert = 1/2$, with the sign being positive (negative) for clockwise (anticlockwise) rotation. The magnitude and direction of the DMI vector and hence the sign of $w_n$, is determined by  a combination of the heavy metal \cite{chen2013tailoring}, ferromagnet interface \cite{LoConte2015} and layer ordering \cite{hrabec2014measuring}. The magnitude of the DMI determines the magnetic ground state of the spin texture and this is characterized by a critical value of the DMI  \cite{thiaville2012dynamics}, $D_{c} = \sqrt{AK_{eff}}/\pi $ ($A$ is the exchange stiffness and $K_{eff}$ is the effective out-of-plane anisotropy). For 0 $< D < D_c$, the spin texture is that of homo-chiral N\'eel DWs. For $D > D_c$,  non-uniform spin textures such as cycloids or skyrmions are shown to have a lower energy than a homogeneous ferromagnetic state. 

Current induced motion of both, homo chiral DWs  \cite{emori2013current,ryu2013chiral,LoConte2015} and skyrmions  \cite{woo2016,Jiang283} have been observed in ultrathin magnetic layers and thus shown to be promising candidates for racetrack memory \cite{parkin2015memory,fert2017magnetic} based applications. One of the main requirements for these applications is the ability to have stable and densely packed memory bits, or in this case densely packed DWs/skyrmions. \citeauthor{zhang_skyrmion-skyrmion_2015}\cite{zhang_skyrmion-skyrmion_2015} recently demonstrated numerically that despite the small size of skyrmions (5-10 nm), a minimum distance of 60 nm is necessary to prevent significant repulsion between them. The distance is approximately equal to the material's DMI mediated helix length \newGVK{(valid for small K$_{eff}$)}, L$_D$ (= $4\pi A/\vert D\vert$). This repulsion is a result of the topological charge \newGVK{(characterized by a two dimensional winding number in case of skyrmions \cite{Braun2012})} of all the DMI stabilized skyrmions. To study this interaction, the problem can be  simplified and studied in a 1D limit. In this case, the skyrmion is simply a pair of chiral DWs, which can be visualized by a vertical cut through the diameter of a 2-dimensional N\'eel skyrmion  \cite{everschor2014real}.  Therefore, a pair of chiral DWs can be used to gain insights into interactions of skyrmions as well as the pure DW interaction. 

The annihilation of domain walls has been proposed\cite{martinez_coupled_2014, hiramatsu2014} and used\cite{benitez2015,del_real_current-induced_2017} to quantify the DMI. However, this has been experimentally studied in thin films where the DWs can assume random orientations depending on the local magnetic parameters. This makes it difficult to study purely the interaction of domain walls. It is therefore important to study locally and with nanometer resolution, small enough to resolve the smallest spin texture (on the exchange length), the DWs and their interaction. The domain walls in films with perpendicular magnetic anisotropy are expected to be only a few (7-15) nm in width. This prevents use of any optical techniques (due to the Rayleigh criteria) and  restricts the option to tunneling and electron/x-ray microscopy imaging techniques such as spin polarized scanning tunneling microscopy \cite{bode2007chiral}, spin polarized low energy electron microscope \cite{chen2013tailoring} and Lorentz transmission electron microscopy \cite{benitez2015}. 

In this work, we comprehensively study the interaction of chiral DWs in a combined theory and experiment effort. We start with an extended analytical model, which allows us to study and modify the dipolar DW repulsion induced by DMI. These results are further developed and used to propose an experimental scheme. This is realized experimentally by performing magneto-transport measurements in Ta$\mid$Co$_{20}$F$_{60}$B$_{20}$$\mid$MgO to check the results of the analytical calculations. To confirm our analytical calculations and compare it to the experimental results, we study it numerically by micromagnetic simulations. This allows us to consider all dipolar interactions, which stabilize the final spin texture and therefore replicate the experimental conditions. 

\section{Analytical model}
Domain walls in continuous film systems with PMA are magnetostatically expected to form Bloch walls. However, the presence of DMI has been shown to facilitate the DWs to adopt a N\'eel character in their ground state \cite{thiaville2012dynamics,bode2007chiral, emori2013current,ryu2013chiral}. The DMI strength, which sets the character of the DW is modelled in terms of an effective magnetic field, called the effective DMI field ($\mu_0H_{DMI}$). This is most apparent at the DW transition and it also  depends on the direction of transition of the magnetization: Up - Down (UD) or Down - Up (DU). The chiral nature of the DMI results in the effective field being opposite in sign for UD and DU DWs. The absolute sign itself is dependent on the layer stack \cite{chen2013tailoring,torrejon_interface_2014}, including the ferromagnet and the heavy metal element whose spin orbit coupling mediates the DMI. The stack used in this study, Ta$\mid$Co$_{20}$F$_{60}$B$_{20}$$\mid$MgO has been previously measured \cite{LoConte2015} to have a positive sign, indicating a right-handed chirality (D$>$0).        

\begin{figure} 
\centering
\includegraphics[width = 1\linewidth]{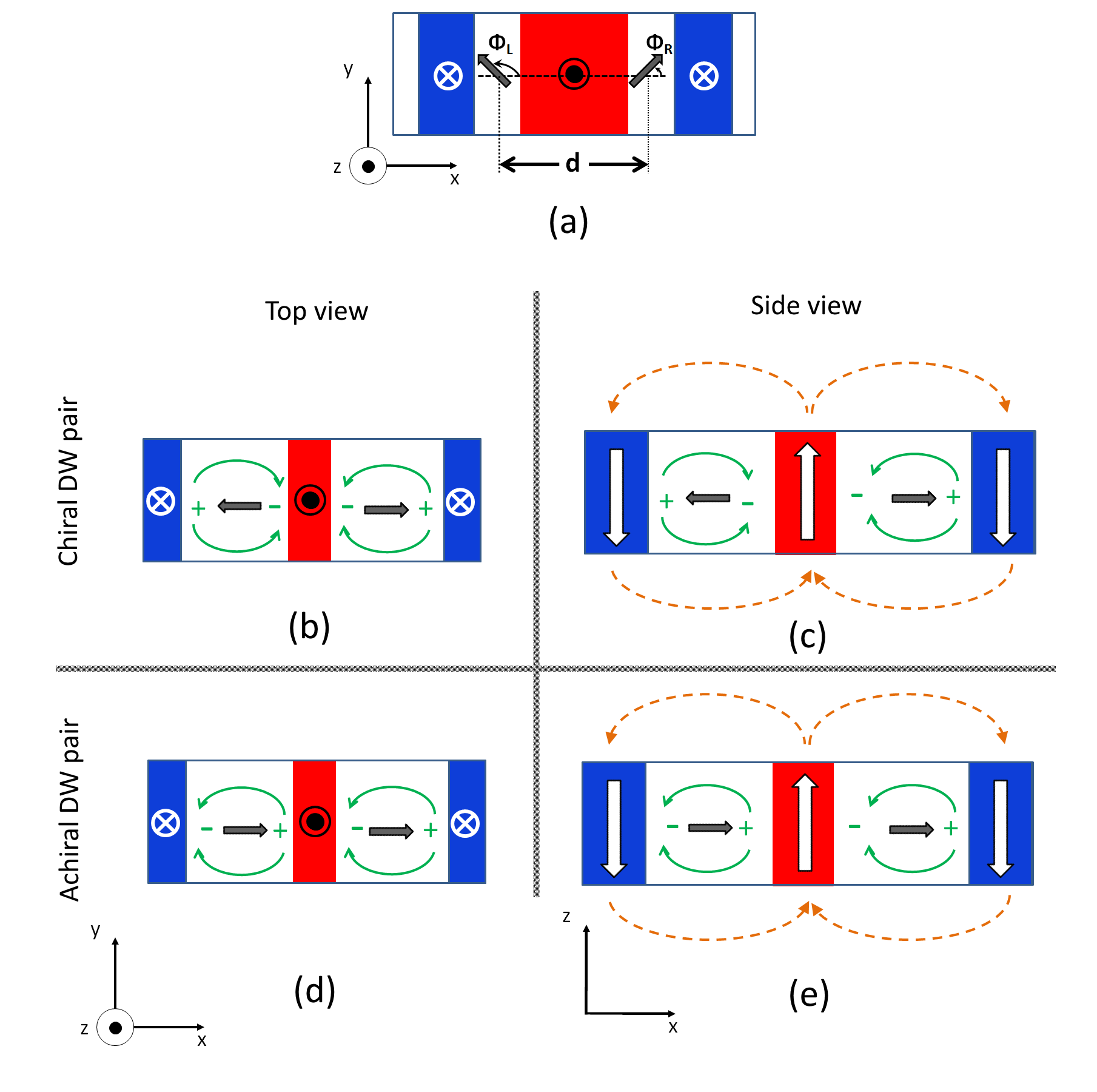}
\caption{ \newGVK{(a) Top view schematic representation of the DW state in the analytical model. $\phi_L$ and $\phi_R$ represent the angles of the internal magnetic moment within the Down-Up (left) and Up-Down(right) domain walls. The angles are considered with respect to the x-axis. (b) \& (d) depict the top view schematic of DW pairs. The green lines indicate the direction of the dipolar fields originating from the in-plane magnetization component present within the DWs for: (b) a chiral DW pair and (d) Achiral DW pair. (c) \& (e) depict the side view schematic of a DW pair. The green line indicates the dipolar fields caused by the in-plane magnetization within the domain wall. The orange dotted lines indicate the dipolar field originating from the domains for: (c) Chiral DW pair and (e) Achiral DW pair.}  
} 
\label{fig:DW_Schematic}
\end{figure}

\newGVK{
In the system under study there are two main interactions influencing the state of the DW pairs (attractive or repulsive interaction between the DWs). These interactions are both related to dipolar effects, and are:
\begin{enumerate}

\item The dipolar interaction between the lateral domains (magnetized down, -z) and the central domain (magnetized up, +z) (see Fig. \ref{fig:DW_Schematic} (c) \& (e)). The dipolar field generated by the lateral domains points up (+z) in the central domain (domain-domain dipolar interaction). Note that this repulsive interaction is independent of the internal magnetic moments within the DWs. This interaction is therefore (lowest order approximation) not influenced by the application of the relatively small in-plane magnetic fields that we study.

\item The dipolar interaction between the magnetic moments within the DWs (see Fig. \ref{fig:DW_Schematic} (b) \& (d)). This interaction depends on the relative direction of the internal magnetic moments (wall-wall dipolar interaction). In the absence of any in-plane  field, the internal magnetic moments are antiparallel (chiral) to each other if the DMI is strong enough. This is due to chiral nature of the DWs imposed by the DMI. In the case of antiparallel magnetic moments, the force experienced by a DW due to the magnetic moment within the other DW is repulsive in nature. However, when the internal magnetic moments are parallel (achiral) to each other the wall-wall interaction becomes attractive. Therefore, this attraction reduces the required out-of-plane field to promote the annihilation of the DW pair. 

%If a longitudinal in-plane field is applied (≠0) with a given polarity (>0, for instance) one of the internal DW moments is supported by this field whereas the other one experiences a torque which tries to rotate the internal magnetic moment along the in-plane field. If this field is strong enough, the internal magnetic moments become parallel to each other, and therefore, the wall-wall interaction becomes attractive. Therefore, this attraction reduces the needed out-of-plane field () to promote the annihilation of the DWs, and this is precisely what our works shows both theoretically and experimentally.
\end{enumerate}
There are several other interactions responsible of the magnetic state in these systems: the exchange interaction, the demagnetizing field of the domains, the magnetic perpendicular anisotropy and the interaction with the external field. However, these interactions, although needed to describe the magnetic state with the two DWs, do not play a significant role on their repulsive or attractive interaction and therefore they are not considered in the simple analytical model discussed henceforth. In particular, the exchange interaction only plays a significant role for domain wall – domain wall interaction for very short distances of the order of the exchange length where the domain wall spin structures become distorted, much shorter than what we study here. The analytical model only considers dipolar interactions from 1. and 2. 

%Therefore the dipolar interaction which is influenced by the application of in-plane magnetic fields is only due to the state of the internal magnetic moments within DWs (which is set initially by the DMI), which is the main focus of this work.

}

We first study the interaction of two neighboring domain walls in the framework of an extended analytical model. Fig. \ref{fig:DW_Schematic} depicts the DWs under study and the variables representing its orientation. $\phi_L$ ($\phi_R$) indicates the angle of the wall magnetization with respect to the x-axis for the left (right) wall. For intermediate strength DMI, the internal magnetization within the walls is expected to be intermediate between Bloch and N\'eel. This results in $\phi_{L,R}$ being intermediate between 0 - 180 \degree

The interaction of two domain walls in close proximity and their dipolar interaction is described by a dipolar force \cite{martinez_coupled_2014} which is \newGVK{generated by the internal magnetic moment of one wall in the center of the other one}:

\begin{widetext}
\begin{equation}
\label{F_dipolar}
 \vec{F}_{dip,L \rightarrow R} = {F}_{dip,L \rightarrow R} \vec{u}_{dip,L \rightarrow R} = \frac{3\mu_0(M_st_{FM} w \Delta)^2}{4\pi d^4} (\sin\phi_L \sin\phi_R - 2\cos\phi_L \cos\phi_R)\vec{u}_{dip,L \rightarrow R}
\end{equation}
\end{widetext}
and the corresponding dipolar field is given by \cite{martinez_coupled_2014}:

\begin{equation}
\label{H_dipolar}
\mu_0H_{dip} (\mu_0H_x, d) = \frac{ F_{dip,L \rightarrow R} }   {2 t_{FM} w M_s} \vec{u_z} 
\end{equation}

 where $\vec{F}_{dip,L \rightarrow R}$ is the  vector representing the dipolar interaction force between domain walls induced by the DMI. $\vec{u}_{dip,L \rightarrow R}$ is the unit vector from left to right. $M_s$ is the saturation magnetization of the ferromagnetic sample. $w$ is the width of the Hall cross, $d$ is the separation distance between the two domain walls, $t_{FM}$ is the thickness of the ferromagnetic layer and $\Delta$ is the DW width. 
 
Eqns. (\ref{F_dipolar}) and (\ref{H_dipolar}) indicate that this interaction can be easily modified by changing the magnetization angle within the DWs. The equilibrium magnetization angle within the DWs under application of an external in-plane field is given by \cite{je_asymmetric_2013}: 
\begin{equation}
\begin{aligned}
\label{angle}
% refer Je, S.-G. et al.  Phys. Rev. B 88, 214401 (2013). and Gross, I. et al. Phys. Rev. B 94, 064413 (2016).
\phi_L = acos (\pi M_s (\mu_0 H_x - \mu_0 H_{DMI})/4 K_D) \\ % H_DMI = \mu_0 H_DMI and is positive for right handed DW (UD). put the sign and magnitude into the equation
\phi_R = acos (\pi M_s (\mu_0 H_x + \mu_0 H_{DMI})/4 K_D)  % H_DMI = \mu_0 H_DMI and is negative for right handed DW (DU). put the sign and magnitude of the in-plane field into the equation
\end{aligned}
\end{equation}

However, under the influence of the in-plane field there is also a change in the DW width and this can be characterized by 
\begin{widetext}
\begin{equation}
\begin{aligned}
\label{DWwidth}
\Delta_{L,R}(\mu_0 H_x, \mu_0 H_y)  = \sqrt{\frac{2A}{2(K_0 + K_D \sin^2 \phi_{L,R}) - \mu_0 M_s \pi (H_x \cos\phi_{L,R} + H_y \sin\phi_{L,R})}}
\end{aligned}  
\end{equation}
\end{widetext}

The previously determined magnetic parameters \cite{LoConte2015} of Ta$\mid$Co$_{20}$F$_{60}$B$_{20}$$\mid$MgO are considered  for the calculations. $M_s = 1.1\times 10^6 Am^{-1}$. The width of the Hall bar, $w$ = 400 nm and exchange stiffness \cite{emori2013current}, $A$  = $10^{-11} Jm^{-1}$. $ \mu_0 H_K$  (400 mT) is the measured saturation field along the magnetic hard-axis and $K_0= \mu_0 H_K M_s/2$ = ($2.2 \times 10^5 J/m^3$) is the effective anisotropy \cite{LoConte2015} and $\Delta_0 = \sqrt{A/K_0}$ = 7 nm is the DW width parameter at equilibrium. $K_D = N_x\mu_0M_s^2/2$ is the DW anisotropy energy density \cite{Tarasenko199819} and $N_x (= t_{FM} ln(2)/(\pi\Delta)$) is the demagnetization factor \cite{Tarasenko199819}. The effective DMI field\cite{LoConte2015} is $\mu_0 H_{DMI}$ = 8 mT.

Using Eqn. (\ref{angle}) - (\ref{DWwidth}) in Eqn. (\ref{F_dipolar}) and Eqn. (\ref{H_dipolar}) we study the change in the dipolar interaction as a function of an additional applied in-plane field ($\mu_0H_x$). \newGVK{
The analytical model derived here considers two free DWs, whose internal magnetic moments are determined by the DMI interaction and the in-plane magnetic field. The force between the DWs depends on the relative distance between the DWs and on the relative orientation of their internal magnetic moments (in the absence of pinning, defects etc., which do not affect the qualitative understanding of realistic experimental scenarios). For systems where the exchange interaction is frustrated a more complex model with additional Hamiltonian terms needs to be employed \cite{leonov_multiply_2015}.
}

\begin{figure} 
\centering
\includegraphics[width = 0.8\linewidth]{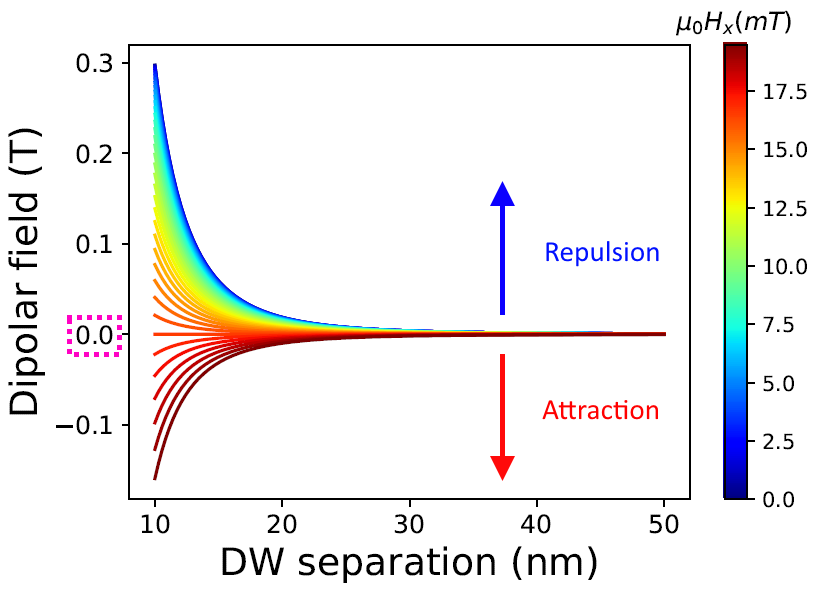}
\caption{Dipolar field \newGVK{(generated by the internal magnetic moment of one wall in the center of the other one)} for various in-plane fields and its dependence on the separation between DWs. The magnitude of the dipolar field directly indicates the external field ($\mu_0 H_z$) needed to produce an attractive force between the domain walls and finally annihilate them. }
\label{fig:DipField_variousIP}
\end{figure}

Fig. \ref{fig:DipField_variousIP} shows the variation of the dipolar field with distance and in-plane fields. We clearly observe that by increasing the magnitude of the magnetic in-plane field, we are able to tune the dipolar interaction between the domain walls. We modify the interaction from a repulsive (positive) to an attractive (negative) interaction at higher magnetic fields. This indicates that the in-plane field is able to tune the N\'eel character of the DW, resulting in neighboring DWs having opposite winding number, +1/2 and -1/2 for high magnetic fields. We notice that the dipolar interaction is strongest when the two domain walls are close to each other. With increasing separation, this interaction reduces and therefore the dipolar repulsion as well. The reduction in the dipolar field with $\mu_0 H_x$ indicates that the field ($\mu_0 H_z$) needed to reduce the dipolar repulsion between the domain walls and annihilate the DWs is reduced.

%\begin{figure}[!ht]

\begin{figure} 
     
       \includegraphics[width = 0.5\linewidth]{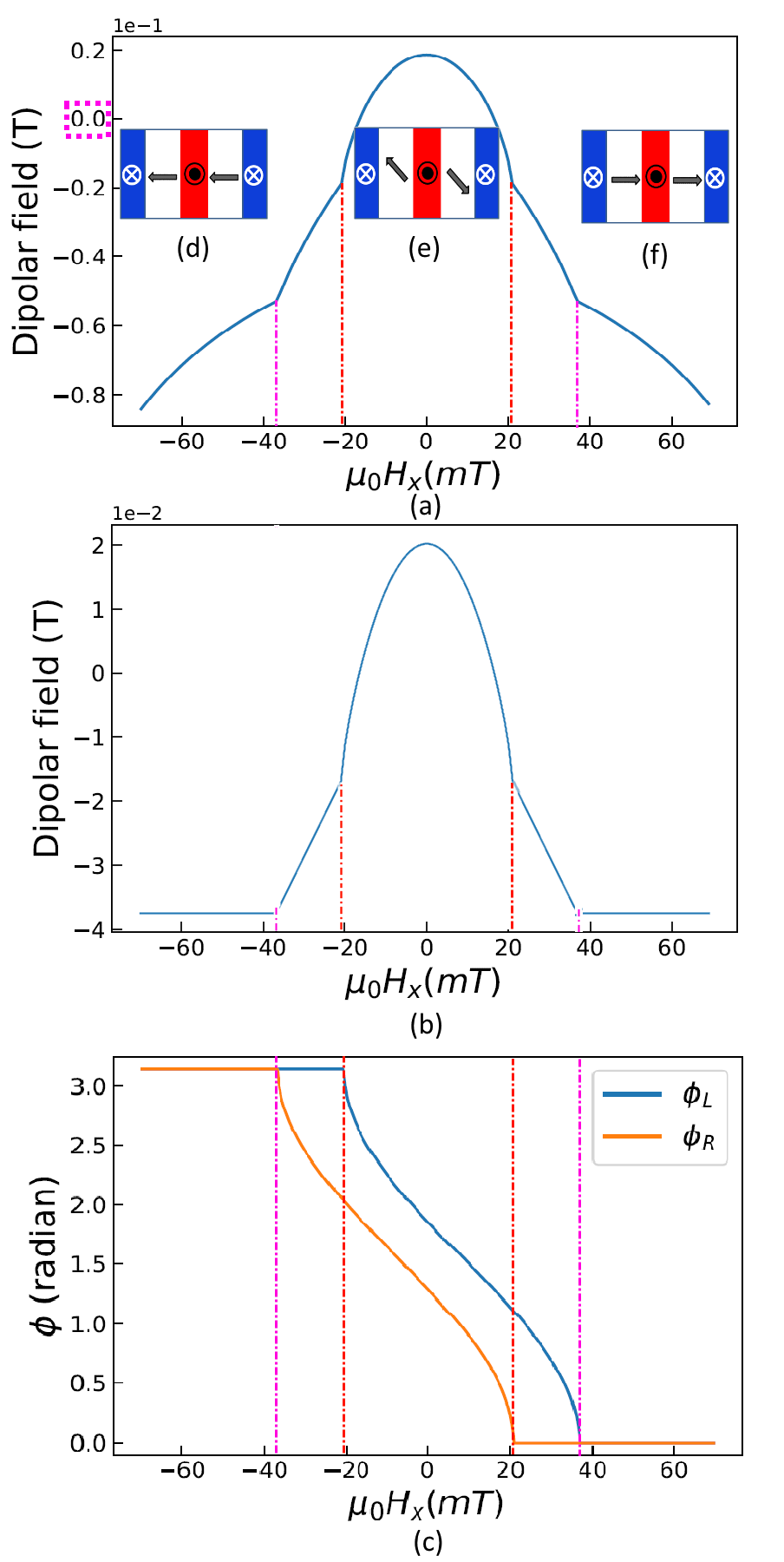}
     
     \caption{(a) \& (b) The dipolar field \newGVK{(generated by the internal magnetic moment of one wall in the center of the other one)} as a function of in-plane applied field when calculated for a fixed separation of DWs (d = 20 nm). The field changes from repulsive (positive) to attractive (negative) under the application of the in-plane field ($\mu_0 H_x$). When variation of the DW width is not considered (as in (b)), there is no change in the dipolar field after the magnetization within both the DWs are saturated in the same direction under application of an in-plane field. (c) Domain wall magnetization angle as a function of $\mu_0H_x$. The magnetization within the wall can be tuned and its chirality changed by the application of an in-plane field. (d)-(f) Top view schematic representation of the DW state at $\mu_0 H_x <$ 36 mT, $\mu_0 H_x$ = 0 and  $\mu_0 H_x >$ 36 mT respectively.}
     \label{fig:DW_Angle_analytical}
   \end{figure}

To gauge the impact of an applied field, we next study the dipolar interaction for a fixed separation of the DWs under different strengths of magnetic fields (see Fig. \ref{fig:DW_Angle_analytical} (a)). There are four main characteristics that we extract  from this analysis: a) We observe that within the limitations of the model, the dipolar repulsion precisely goes to zero when the external magnetic field compensates for the effective DMI field (2$\mu_0H_{DMI}$ = 16 mT). b) The first - abrupt change occurs at around 20 mT (see also Fig. \ref{fig:DW_Angle_analytical} (c)) and this takes place when one of the walls is in a completely N\'eel state ($\phi_{L\:or\:R} = 0 \degree $ or $ 180\degree $). c) The second - kink occurs at around 36 mT  (which is = 20 mT + 2$\mu_0 H_{DMI}$, see also Fig. \ref{fig:DW_Angle_analytical} (c)). This occurs when both walls are completely in the N\'eel state ($\phi_{L\:and\:R} = 0 \degree $ or $ 180\degree $). d) The dipolar interaction continues being attractive for larger fields, $\vert \mu_0 H_x\vert >$ 36 mT, purely due to the widening of the domain wall widths in the system (see Fig. \ref{fig:DW_Angle_analytical} (a)). In the absence of increase of DW widths, the dipolar interaction will be saturated at $\vert \mu_0 H_x\vert >$ 36 mT (see Fig. \ref{fig:DW_Angle_analytical} (b)). These results show how the various intertwined variables need to be considered to precisely determine the dipolar interaction of the domain wall pairs and eventually of skyrmions.

\section{Experimental results}
To study the interaction experimentally, we propose a experimental scheme to probe the predictions of the analytical calculations. \newGVK{However, in comparison to the free DWs that we presented in the analytical model, here we study the DW interaction in a Hall-cross device structure with two DWs pinned at the Hall-cross entrance (thus can also include lengthening/elongation of the DWs).} We perform the experimental study in a system of Ta$\mid$Co$_{20}$F$_{60}$B$_{20}$$\mid$MgO where it has been shown \cite{LoConte2015} that the N\'eel character of the domain wall in this system can be tuned by application of an in-plane magnetic field ($\mu_0H_x$). 

\begin{figure} 
     \subfloat[\label{fig:Exp_HallBar}]{%
       \includegraphics[width= 0.5\textwidth]{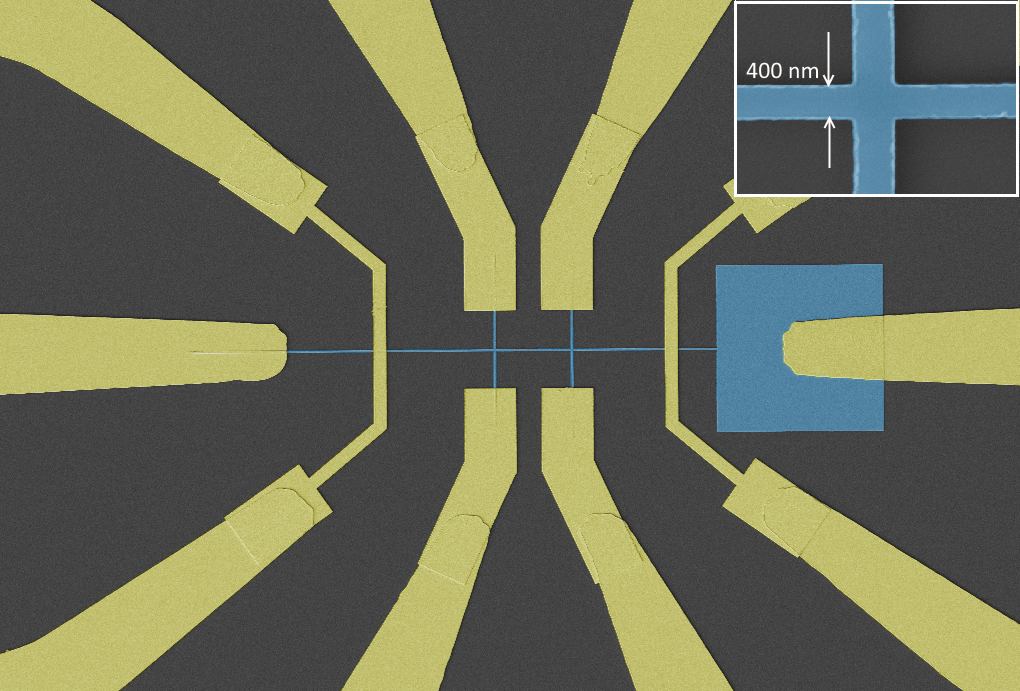}
     }
     \hfill
     \subfloat[\label{fig:SysInitial}]{%
       \includegraphics[width= 0.5\textwidth]{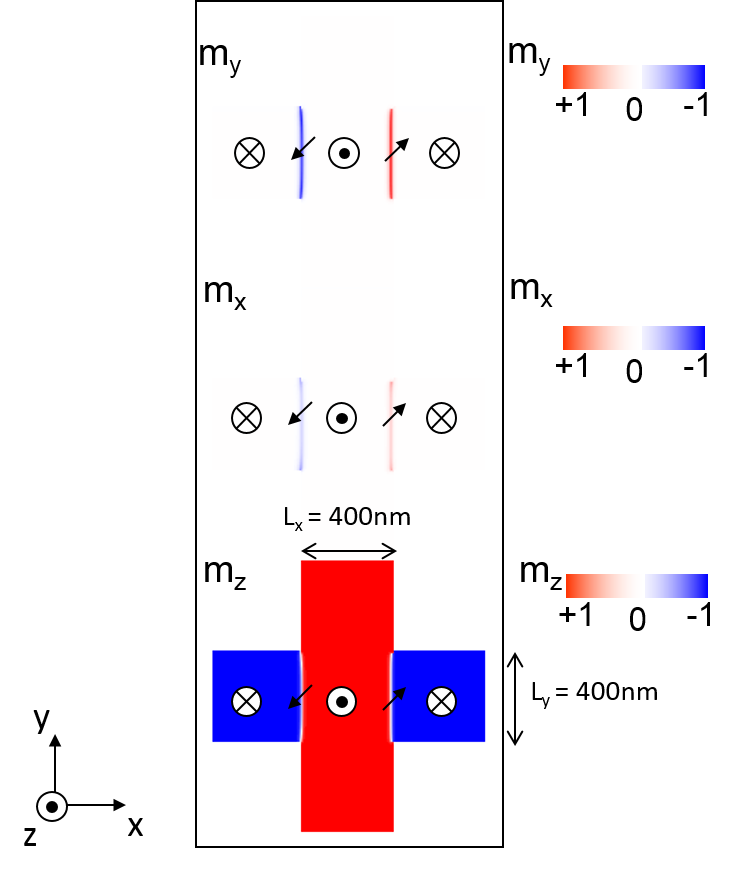}
     }
     \caption{ (a) Scanning electron microscopy image of the experimental device structure. It consists of a 400 nm wide Hall cross (blue) and  contact pads (gold). The inset shows a zoomed in picture of the Hall cross. (b) Initialized state with two domain walls placed at the exit of the Hall cross, as used in the micromagnetic simulations. $m_x$, $m_y$ and $m_z$ show the corresponding magnetization states as indicated by the colour bar.}
     \label{fig:Initialimages}
   \end{figure}

The sample of Ta(5)$\mid$Co$_{20}$F$_{60}$B$_{20}(1)$$\mid$MgO(2)$\mid$Ta(5) (all thicknesses in nm) is deposited by sputtering (using a Singulus TIMARIS/ROTARIS tool). The samples are grown on a thermally oxidized silicon wafer and are annealed post-deposition at 300 \degree C under vacuum for 2 hours. This is done to achieve a large perpendicular magnetic anisotropy. To perform the experiments, Hall bars ($w$ = 400 nm) are patterned using e-beam lithography and argon-ion milling (see Fig. \ref{fig:Exp_HallBar}).

Transport measurements are performed on a set-up consisting of a 3D vector magnet. This allows one to precisely align the magnetic field along any direction of the sample. Experimentally we study the interaction of domain walls by making use of the anomalous Hall effect (AHE) \cite{hurd1972hall}, which provides high sensitivity for probing the DWs.

The sample is initially saturated in the down magnetization state by applying a field, $\mu_0H_z$ = - 2 T. The AHE is measured across the Hall bar, which is aligned along the y-axis. To precisely study the modification of the domain wall under in-plane fields, it is important to align the wire along the  x-axis and the bar along the y-axis. This is done by measuring the planar Hall effect while performing an angular scan of the magnetic field (0.9 T) along the x-y plane. By fitting the anisotropic magneto resistance we are able to quantify and precisely align the magnetic field with respect to the sample orientation. The change of the AHE signal is directly related to the change of the out-of-plane magnetization in the cross and hence this sensitively detects the change in the magnetic state. While continuously measuring the AHE we gradually increase the magnitude of the out-of-plane field (- $\mu_0H_z$). The percentage change in the AHE reflects the change in the area of magnetization in the Hall bar that is pointing along the + and - z-direction. 

To create the initial state we make use of spin orbit torque (SOT) driven switching \cite{Liu555,garello2013symmetry} . We initially saturate the magnetization in the sample in the down-state by applying a field along - $\mu_0H_z$ (2 T). We then apply a current pulse of 100 ns duration and $\approx$ $10^{11}$ A/$m^2$ current density along +y, during the application of an in-plane field ($\mu_0H_y$ = + 50 mT). This results in complete switching of the magnetization in the Hall bar, which is reflected in the AHE voltage. This leads to creation of two DWs at the corners of the cross entrance: a Down-Up (DU) wall on the left side and a Up-DW (UD) wall at the right side. The DMI present in the system induces a handedness to the DWs, resulting in a pair of homo-chiral walls separated by 400 nm (width of the Hall cross).

\begin{figure} 
     
       \includegraphics[width= 1\textwidth]{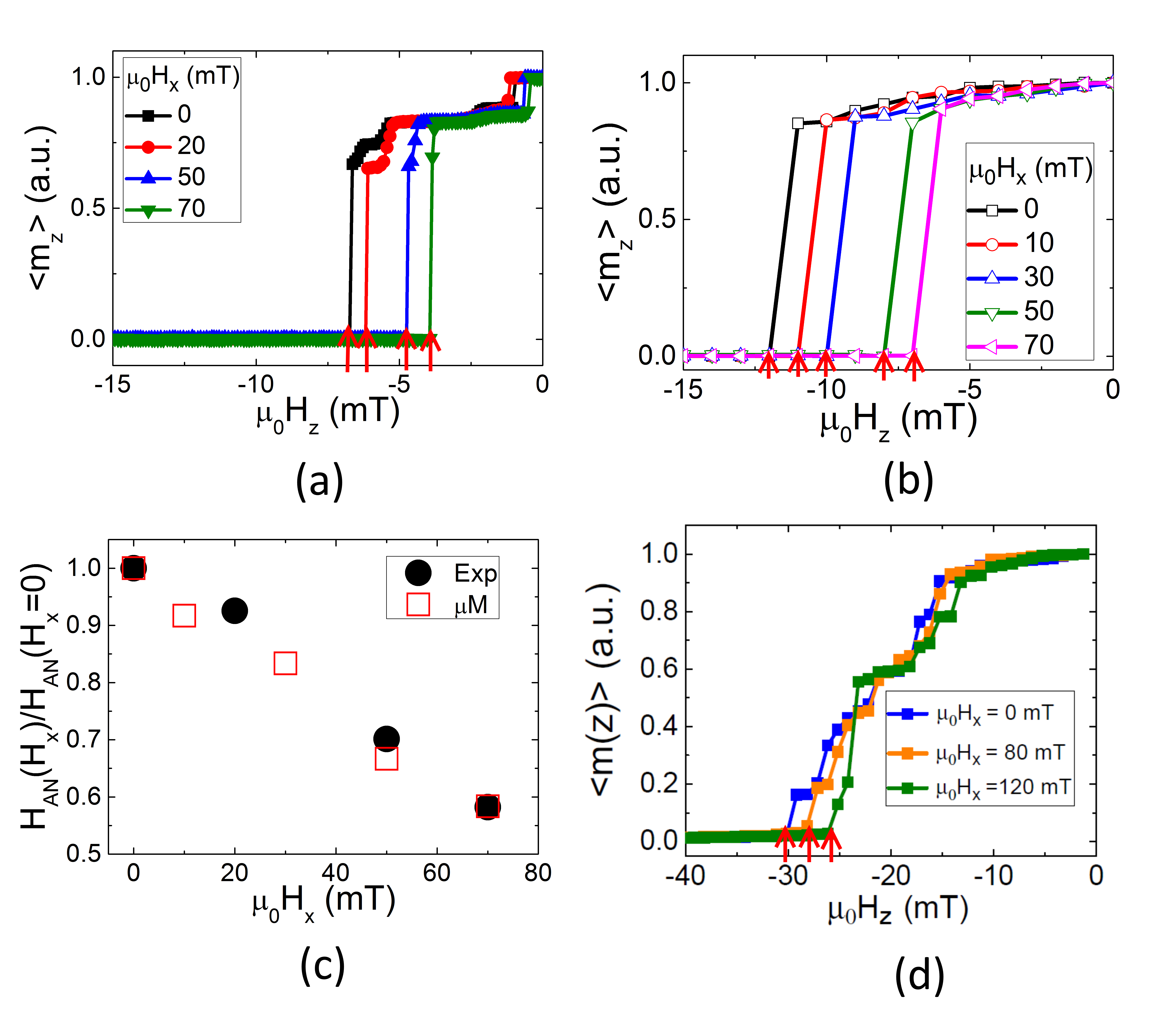}

     \caption{ \newGVK{Changes in the magnetization in the central part of the Hall bar under the application  of  magnetic fields by (a) experimentally measured anomalous Hall voltage and (b) micromagnetic simulations, in a system of Ta$\mid$Co$_{20}$F$_{60}$B$_{20}$$\mid$MgO (D = +0.06 $mJ/m^2$). The normalized anomalous Hall voltage plots indicate the change in the z-component of magnetization and the terminal field reflects the annihilation of the domain walls. This is studied under the application  of  magnetic fields. While $\mu_0H_x$ tunes the N\'eel character of the domain walls,  $\mu_0H_z$ drives the motion of the domain walls. (c) Normalized annihilation field (derived from (a) and (b)) is plotted as a function of applied in-plane field ($\mu_0H_x$), shows agreement between the micromagnetic simulations (red square) and experimental (black filled circles) results. (d) control experimental measurement performed in a system of Pt$\mid$Co$\mid$AlO$_x$ \cite{LoConte_2017} (D = +0.44 $mJ/m^2$ ). The annihilation field in this system is higher and is similarly influenced by the application of in-plane magnetic fields. The red arrows indicate the annihilation field (H$_{AN}$).}
     }
     \label{fig:Annihilation_AHE}
   \end{figure}

\begin{figure}

\centering
\includegraphics[width = 1\linewidth]{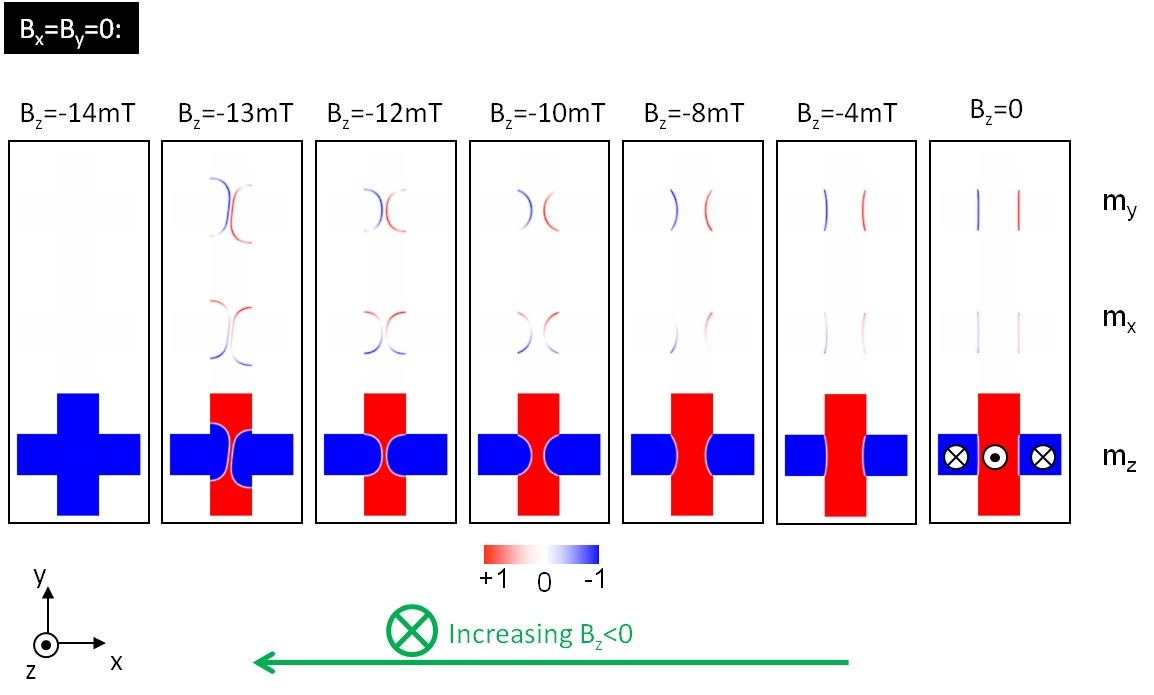}
\caption{Interaction and annihilation of domain walls under the application of $\mu_0H_z$. The images show the snapshot of the magnetization profile within the wall and its evolution under the application of external magnetic field, from micromagnetic simulations for a perfect sample (without defects) and at zero temperature (T=0). Note that results in Fig. 5(b) are from simulations which were computed in the presence of defects, and at room temperature. } 
\label{fig:MicroMagn_Annihi_0}
\end{figure}

We have already seen that the impact of the dipolar repulsion due to the DMI can be modified. This is done by applying an in-plane field (along x-axis), which changes the N\'eel character of both the walls. The in-plane field is able to modify the walls to eventually have different winding number and thus result in a reduction of their dipolar repulsion. Due to the application of a magnetic field, there is a change in the Zeeman energy of the domains. This results in an increase in the size of the domains aligned along the magnetic field. Therefore under the application of -$\mu_0H_z$ the central domain shrinks while the two domains in the extrema expand (see Fig.\ref{fig:MicroMagn_Annihi_0}). \newGVK{Here the extrema domains refer to the domains which bound
the central domain.} This results in the motion of the two domain walls towards each other. When the externally applied out-of-plane field is strong enough to compensate for the dipolar repulsion induced by the two extreme domains and the DMI (parameterized by D), the two domain walls finally annihilate, and this process switches the complete Hall bar (reflecting a down magnetization in the Hall bar as well) as shown in Fig. \ref{fig:MicroMagn_Annihi_0}. This terminal field is reflective of both the magnitude of the dipolar repulsion caused by the DMI and the two extreme domains. The slight reduction in the AHE signal at $\approx$ -1.5 mT \newGVK{(see Fig. \ref{fig:Annihilation_AHE} (a))} can be ascribed to local structural defects which result in domain wall pinning and this is found to be structure/device dependent. In our case this defect mediated pinning field is sufficiently low to observe the terminal field for domain wall annihilation and the influence of in-plane magnetic fields. 

 The same experiment was repeated, but now in the presence of in-plane fields. We observe (see Fig. \ref{fig:Annihilation_AHE} (a)) that, the annihilation field decreases as the magnitude of the in-plane field increases. This indicates that under the application of an in-plane field  it is possible to break the fixed chirality present in the system. This in turn modifies the dipolar repulsion  stabilizing the winding pair. We thus find a dependence of the annihilation field on the in-plane applied field, which is in agreement with the results from the energetics calculated in our analytical model.   

\newGVK{In order to further confirm our experimental observation we repeat the experimental procedure on a system of Pt$\mid$Co$\mid$AlO$_x$ \cite{LoConte_2017}, with $t_{FM}$
 = 1 nm,
$K_{eff}$ = 0.95 $\times$ 10$^6$ J/m$^3$
$M_s$= 1423 kA/m
and A = 1.6 $\times$ 10$^{−11}$ J/m \cite{thiaville2012dynamics}
. As the DMI parameter (D) increases, the chiral configuration (anti-parallel internal magnetic moments within the DW pair) is supported, and consequently higher in-plane field is required to fully reverse from this state to the one with parallel internal magnetic moments. Therefore, the annihilation field H$_{AN}$ (defined as the out-of-plane field needed to annihilate the DWs) increases with D \cite{martinez_coupled_2014,hiramatsu2014,del_real_current-induced_2017}. From Fig. \ref{fig:Annihilation_AHE} (d) we observe that the annihilation field for Pt$\mid$Co$\mid$AlO$_x$ (D = +0.44 $mJ/m^2$ ) \cite{LoConte_2017} is indeed higher than for Ta$\mid$Co$_{20}$F$_{60}$B$_{20}$$\mid$MgO (D = +0.06 $mJ/m^2$) \cite{LoConte2015}. However, it should be noted it is not possible to experimentally extract a quantitative correlation between the annihilation field and the DMI in two different samples, as there is simultaneous variation of magnetic parameters and material parameters (grain size, defects etc.,).
}

\section{Micromagnetic simulations}
In order to check the analytical predictions and to better develop our interpretation of the experimental observations, full micromagnetic simulations have been performed. The micromagnetic simulations are performed for a realistic experimental set-up, including all magnetostatic interactions-including dipolar contributions from the extrema domains\cite{kim_control_2011}.  The dimensions of the Hall cross are the same as used in the experimental study. The computational region consists of two orthogonal ferromagnetic strips forming the Hall cross, where each of them has a cross section of w $\times$ t$_{FM}$ = 400 nm $\times$  1 nm (see Fig. 5). %The length of each orthogonal ferromagnetic strip is, l = 1200 nm. 
\newGVK{A computational square region of l = 2048 nm centred in the cross Hall is evaluated; both arms of the cross Hall have the same transverse cross section (w $\times$ t$_{FM}$).} Typical experimental material parameters of Ta$\mid$Co$_{20}$F$_{60}$B$_{20}$$\mid$MgO are adopted \citep{LoConte2015} for the micromagnetic modeling, and are the same as used for the analytical study. \newGVK{The computational region was discretized using a 2D finite difference scheme with cells of $\Delta$x = $\Delta$y= 4 nm and $\Delta$z = 1 nm. An initial state with the 2 DWs initially pinned at the two lateral sides of the cross Hall was considered. The equilibrium states of the magnetization under both out-of-plane ($B_z$) and longitudinal in-plane ($B_x$) fields were computed by two methods using Mumax3 software: i) by minimizing the total energy of the system, and ii) by solving the LLG eq. using a Gilbert damping of, $\alpha$ = 0.5. Both methods returned similar results of decrease of annihilation field on application of in-plane field.
%Micromagnetic simulations were computed under both ideal and realistic conditions. For the ideal case, perfect samples at zero temperature were evaluated, and the equilibrium states of the under both static out-of-plane and in-plane fields we computed as describe above: either i) by minimizing the total energy of the system, and ii) by solving the LLG eq. using a Gilbert damping of $\alpha$ = 0.5. Both methods resulted in similar results. 
Additionally, micromagnetic simulations were computed under more realistic conditions. In order to take into account the effects of disorder due to imperfections and defects in a more realistic way, we assume that the easy axis anisotropy direction ( $\vec{u}_k= \vec{u}_k(\vec{r}_g)$) is distributed among a length scale defined by a grain size \cite{del_real_current-induced_2017}. The average size of the grains is 10 nm. Despite the fact that the direction of the uniaxial anisotropy of each grain is mainly directed along the perpendicular direction (z-axis), a small in-plane component lower than 10 \% is randomly generated over the grains. Although other ways to account for imperfections could be adopted, we selected this one based on previous studies \cite{del_real_current-induced_2017}, which properly describe other experimental observations in related systems. In this realistic modeling, the effect of thermal fluctuations was also considered. Thermal fluctuations at room temperature are taken into account by adding a random stochastic thermal field to the deterministic effective field, so the deterministic LLG eq. becomes an stochastic eq., which is also numerically solved \cite{del_real_current-induced_2017}. 
}

The initial equilibrium state consists of two DWs initially pinned at the two lateral sides of the cross. The initialized state, along with the magnetization components ($m_x, m_y, m_z$) in the domain wall is shown in Fig. \ref{fig:SysInitial}.  Note that, due to the relative small magnitude of the DMI in these samples, the internal magnetic moment within each DW is pointing along an intermediate direction within the xy plane and therefore the DWs depict an intermediate configuration between Bloch and N\'eel states.

Starting from this state, and to evaluate the minimum field at which the DWs annihilate, a series of decreasing out-of-plane fields ($-B_z=-\mu_0 H_z:0$ upto -20 mT) were applied and the new equilibrium state was numerically computed using Mumax3 \cite{Vansteenkiste_MuMax_2014}. %Simulations were carried out by looking to the equilibrium state for each external field. 
In the absence of in-plane field $B_x=\mu_0 H_x=0$), the annihilation field is $-B_z\approx-14 mT$ (see Fig. \ref{fig:Annihilation_AHE} (a) and \ref{fig:Annihilation_AHE} (b)). However, the magnitude of annihilation field ($ \mid B_z \mid$) decreases as the magnitude of the in-plane field ($B_x$) increases and this is in good agreement with the experimental observations (see Fig. \ref{fig:Annihilation_AHE} (a) ) and with the analytical predictions (Eqn. \ref{F_dipolar}-\ref{H_dipolar}), which predict a decrease in the repulsive force between the DWs as their internal magnetic moment become parallel due to strong in-plane fields.
\newGVK{The micromagnetic results were computed under the mentioned conditions. These results show a good agreement with the experimental data. In particular, it has to be noticed that the experimental and micromagnetic simulation results demonstrate agreement when the normalized annihilation field is plotted as function of the in-plane field (see Fig. \ref{fig:Annihilation_AHE} (d)), which clearly indicates the validity of our description.}
%However, it should be noted that the absolute annihilation field has different values for the simulations and the experiment. This is attributed to thermal activation, which is not taken into account in the modeling. 

\section{Conclusion}
 We have shown that the internal magnetic moments present in neighboring homo-chiral DWs generate an additional dipolar repulsive force. This force is observed to be the dominant interaction when in close ($<$ 15nm) proximity (see Fig. \ref{fig:DipField_variousIP} ). We learn from the analytical calculations that the interaction between two domain walls is non-linear as a function of their separation and changes drastically when they are in close proximity. This prevents conventional imaging techniques from accurately determining their interaction/structure and therefore requires electrical measurement schemes.
 
Additionally, the DMI extracted by using annihilation based experimental schemes can lead to an overestimation of the DMI value. This is because, the domains by themselves interact repulsively when in close proximity\cite{kim_control_2011,martinez_coupled_2014,del_real_current-induced_2017}. Therefore it is important to observe the effect of the annihilation field under in-plane fields, where only the spin texture induced interaction is modified. 

In this work, we developed an extended analytical model and showed that the DMI which induces a homo-chiral nature to the DWs, contributes additionally to the dipolar repulsion between neighboring walls. We show that we can tailor the dipolar interactions by applying an external magnetic field that changes this interaction from a repulsive to an attractive interaction. This is done by changing the sign of the winding number in one of the DWs. This highlights the importance of the chiral nature of the interaction driven by the DMI. While a higher DMI increases the robustness against DW collision for higher DW density devices, this also puts a limit on the maximum packing density possible  before repulsive interaction dominates.  

Finally, we probed the interaction experimentally in a magneto-transport measurement. We determined the annihilation fields experimentally in a system of Ta$\mid$Co$_{20}$F$_{60}$B$_{20}$$\mid$MgO, which has a finite DMI. To understand the experiments and check the analytical calculations, we performed micromagnetic simulations which included all the dipolar interactions present in the experiment. We found that the results of the numerical calculations agree with the experimental observations, showing how the interaction results from the interplay of the spin textures governed by the applied fields and the DMI. 

\begin{acknowledgments}
We acknowledge support by EU
(MAGWIRE, Project No. FP7-ICT-2009-5; Marie Curie ITN WALL, FP7-PEOPLE-2013-ITN 608031) and the DFG (TRR 173). The work by E. M. was additionally supported by Projects No. MAT2014-52477-C5-4-P and MAT2017-87072-C4-1-P from the Spanish government, and Project No. SA090U16 from the Junta de Castilla y Leon.
\end{acknowledgments}
%\pagebreak
%\bibliography{Bibliograph_Annihi}

\begin{thebibliography}{31}%
\makeatletter
\providecommand \@ifxundefined [1]{%
 \@ifx{#1\undefined}
}%
\providecommand \@ifnum [1]{%
 \ifnum #1\expandafter \@firstoftwo
 \else \expandafter \@secondoftwo
 \fi
}%
\providecommand \@ifx [1]{%
 \ifx #1\expandafter \@firstoftwo
 \else \expandafter \@secondoftwo
 \fi
}%
\providecommand \natexlab [1]{#1}%
\providecommand \enquote  [1]{``#1''}%
\providecommand \bibnamefont  [1]{#1}%
\providecommand \bibfnamefont [1]{#1}%
\providecommand \citenamefont [1]{#1}%
\providecommand \href@noop [0]{\@secondoftwo}%
\providecommand \href [0]{\begingroup \@sanitize@url \@href}%
\providecommand \@href[1]{\@@startlink{#1}\@@href}%
\providecommand \@@href[1]{\endgroup#1\@@endlink}%
\providecommand \@sanitize@url [0]{\catcode `\\12\catcode `\$12\catcode
  `\&12\catcode `\#12\catcode `\^12\catcode `\_12\catcode `\%12\relax}%
\providecommand \@@startlink[1]{}%
\providecommand \@@endlink[0]{}%
\providecommand \url  [0]{\begingroup\@sanitize@url \@url }%
\providecommand \@url [1]{\endgroup\@href {#1}{\urlprefix }}%
\providecommand \urlprefix  [0]{URL }%
\providecommand \Eprint [0]{\href }%
\providecommand \doibase [0]{http://dx.doi.org/}%
\providecommand \selectlanguage [0]{\@gobble}%
\providecommand \bibinfo  [0]{\@secondoftwo}%
\providecommand \bibfield  [0]{\@secondoftwo}%
\providecommand \translation [1]{[#1]}%
\providecommand \BibitemOpen [0]{}%
\providecommand \bibitemStop [0]{}%
\providecommand \bibitemNoStop [0]{.\EOS\space}%
\providecommand \EOS [0]{\spacefactor3000\relax}%
\providecommand \BibitemShut  [1]{\csname bibitem#1\endcsname}%
\let\auto@bib@innerbib\@empty
%</preamble>
\bibitem [{\citenamefont {Parkin}\ and\ \citenamefont
  {Yang}(2015)}]{parkin2015memory}%
  \BibitemOpen
  \bibfield  {author} {\bibinfo {author} {\bibfnamefont {S.}~\bibnamefont
  {Parkin}}\ and\ \bibinfo {author} {\bibfnamefont {S.-H.}\ \bibnamefont
  {Yang}},\ }\href@noop {} {\bibfield  {journal} {\bibinfo  {journal} {Nat.
  Nanotechnol.}\ }\textbf {\bibinfo {volume} {10}},\ \bibinfo {pages} {195}
  (\bibinfo {year} {2015})}\BibitemShut {NoStop}%
\bibitem [{\citenamefont {Fert}\ \emph {et~al.}(2017)\citenamefont {Fert},
  \citenamefont {Reyren},\ and\ \citenamefont {Cros}}]{fert2017magnetic}%
  \BibitemOpen
  \bibfield  {author} {\bibinfo {author} {\bibfnamefont {A.}~\bibnamefont
  {Fert}}, \bibinfo {author} {\bibfnamefont {N.}~\bibnamefont {Reyren}}, \ and\
  \bibinfo {author} {\bibfnamefont {V.}~\bibnamefont {Cros}},\ }\href@noop {}
  {\bibfield  {journal} {\bibinfo  {journal} {Nat. Rev. Mater.}\ }\textbf
  {\bibinfo {volume} {2}},\ \bibinfo {pages} {17031} (\bibinfo {year}
  {2017})}\BibitemShut {NoStop}%
\bibitem [{\citenamefont {Dzyaloshinsky}(1958)}]{DZYALOSHINSKY1958241}%
  \BibitemOpen
  \bibfield  {author} {\bibinfo {author} {\bibfnamefont {I.}~\bibnamefont
  {Dzyaloshinsky}},\ }\href {\doibase
  http://dx.doi.org/10.1016/0022-3697(58)90076-3} {\bibfield  {journal}
  {\bibinfo  {journal} {J. Phys. Chem. Solids}\ }\textbf {\bibinfo {volume}
  {4}},\ \bibinfo {pages} {241} (\bibinfo {year} {1958})}\BibitemShut {NoStop}%
\bibitem [{\citenamefont {Moriya}(1960)}]{Moriya}%
  \BibitemOpen
  \bibfield  {author} {\bibinfo {author} {\bibfnamefont {T.}~\bibnamefont
  {Moriya}},\ }\href {\doibase 10.1103/PhysRev.120.91} {\bibfield  {journal}
  {\bibinfo  {journal} {Phys. Rev.}\ }\textbf {\bibinfo {volume} {120}},\
  \bibinfo {pages} {91} (\bibinfo {year} {1960})}\BibitemShut {NoStop}%
\bibitem [{\citenamefont {Fert}(1991)}]{fert1990}%
  \BibitemOpen
  \bibfield  {author} {\bibinfo {author} {\bibfnamefont {A.}~\bibnamefont
  {Fert}},\ }in\ \href {\doibase 10.4028/www.scientific.net/MSF.59-60.439}
  {\emph {\bibinfo {booktitle} {Metallic Multilayers}}},\ \bibinfo {series}
  {Materials Science Forum}, Vol.~\bibinfo {volume} {59}\ (\bibinfo
  {publisher} {Trans Tech Publications},\ \bibinfo {year} {1991})\ p.\ \bibinfo
  {pages} {439}\BibitemShut {NoStop}%
\bibitem [{\citenamefont {Thiaville}\ \emph {et~al.}(2012)\citenamefont
  {Thiaville}, \citenamefont {Rohart}, \citenamefont {Ju{\'e}}, \citenamefont
  {Cros},\ and\ \citenamefont {Fert}}]{thiaville2012dynamics}%
  \BibitemOpen
  \bibfield  {author} {\bibinfo {author} {\bibfnamefont {A.}~\bibnamefont
  {Thiaville}}, \bibinfo {author} {\bibfnamefont {S.}~\bibnamefont {Rohart}},
  \bibinfo {author} {\bibfnamefont {{\'E}.}~\bibnamefont {Ju{\'e}}}, \bibinfo
  {author} {\bibfnamefont {V.}~\bibnamefont {Cros}}, \ and\ \bibinfo {author}
  {\bibfnamefont {A.}~\bibnamefont {Fert}},\ }\href@noop {} {\bibfield
  {journal} {\bibinfo  {journal} {EPL}\ }\textbf {\bibinfo {volume} {100}},\
  \bibinfo {pages} {57002} (\bibinfo {year} {2012})}\BibitemShut {NoStop}%
\bibitem [{\citenamefont {Bode}\ \emph {et~al.}(2007)\citenamefont {Bode},
  \citenamefont {Heide}, \citenamefont {Von~Bergmann}, \citenamefont
  {Ferriani}, \citenamefont {Heinze}, \citenamefont {Bihlmayer}, \citenamefont
  {Kubetzka}, \citenamefont {Pietzsch}, \citenamefont {Bl{\"u}gel},\ and\
  \citenamefont {Wiesendanger}}]{bode2007chiral}%
  \BibitemOpen
  \bibfield  {author} {\bibinfo {author} {\bibfnamefont {M.}~\bibnamefont
  {Bode}}, \bibinfo {author} {\bibfnamefont {M.}~\bibnamefont {Heide}},
  \bibinfo {author} {\bibfnamefont {K.}~\bibnamefont {Von~Bergmann}}, \bibinfo
  {author} {\bibfnamefont {P.}~\bibnamefont {Ferriani}}, \bibinfo {author}
  {\bibfnamefont {S.}~\bibnamefont {Heinze}}, \bibinfo {author} {\bibfnamefont
  {G.}~\bibnamefont {Bihlmayer}}, \bibinfo {author} {\bibfnamefont
  {A.}~\bibnamefont {Kubetzka}}, \bibinfo {author} {\bibfnamefont
  {O.}~\bibnamefont {Pietzsch}}, \bibinfo {author} {\bibfnamefont
  {S.}~\bibnamefont {Bl{\"u}gel}}, \ and\ \bibinfo {author} {\bibfnamefont
  {R.}~\bibnamefont {Wiesendanger}},\ }\href@noop {} {\bibfield  {journal}
  {\bibinfo  {journal} {Nature}\ }\textbf {\bibinfo {volume} {447}},\ \bibinfo
  {pages} {190} (\bibinfo {year} {2007})}\BibitemShut {NoStop}%
\bibitem [{\citenamefont {Braun}(2012)}]{Braun2012}%
  \BibitemOpen
  \bibfield  {author} {\bibinfo {author} {\bibfnamefont {H.-B.}\ \bibnamefont
  {Braun}},\ }\href@noop {} {\bibfield  {journal} {\bibinfo  {journal} {Adv.
  Phys.}\ }\textbf {\bibinfo {volume} {61}},\ \bibinfo {pages} {1} (\bibinfo
  {year} {2012})}\BibitemShut {NoStop}%
\bibitem [{\citenamefont {Chen}\ \emph {et~al.}(2013)\citenamefont {Chen},
  \citenamefont {Ma}, \citenamefont {N'Diaye}, \citenamefont {Kwon},
  \citenamefont {Won}, \citenamefont {Wu},\ and\ \citenamefont
  {Schmid}}]{chen2013tailoring}%
  \BibitemOpen
  \bibfield  {author} {\bibinfo {author} {\bibfnamefont {G.}~\bibnamefont
  {Chen}}, \bibinfo {author} {\bibfnamefont {T.}~\bibnamefont {Ma}}, \bibinfo
  {author} {\bibfnamefont {A.}~\bibnamefont {N'Diaye}}, \bibinfo {author}
  {\bibfnamefont {H.}~\bibnamefont {Kwon}}, \bibinfo {author} {\bibfnamefont
  {C.}~\bibnamefont {Won}}, \bibinfo {author} {\bibfnamefont {Y.}~\bibnamefont
  {Wu}}, \ and\ \bibinfo {author} {\bibfnamefont {A.}~\bibnamefont {Schmid}},\
  }\href@noop {} {\bibfield  {journal} {\bibinfo  {journal} {Nat. Commun.}\
  }\textbf {\bibinfo {volume} {4}},\ \bibinfo {pages} {2671} (\bibinfo {year}
  {2013})}\BibitemShut {NoStop}%
\bibitem [{\citenamefont {Lo~Conte}\ \emph {et~al.}(2015)\citenamefont
  {Lo~Conte}, \citenamefont {Martinez}, \citenamefont {Hrabec}, \citenamefont
  {Lamperti}, \citenamefont {Schulz}, \citenamefont {Nasi}, \citenamefont
  {Lazzarini}, \citenamefont {Mantovan}, \citenamefont {Maccherozzi},
  \citenamefont {Dhesi}, \citenamefont {Ocker}, \citenamefont {Marrows},
  \citenamefont {Moore},\ and\ \citenamefont {Kl\"aui}}]{LoConte2015}%
  \BibitemOpen
  \bibfield  {author} {\bibinfo {author} {\bibfnamefont {R.}~\bibnamefont
  {Lo~Conte}}, \bibinfo {author} {\bibfnamefont {E.}~\bibnamefont {Martinez}},
  \bibinfo {author} {\bibfnamefont {A.}~\bibnamefont {Hrabec}}, \bibinfo
  {author} {\bibfnamefont {A.}~\bibnamefont {Lamperti}}, \bibinfo {author}
  {\bibfnamefont {T.}~\bibnamefont {Schulz}}, \bibinfo {author} {\bibfnamefont
  {L.}~\bibnamefont {Nasi}}, \bibinfo {author} {\bibfnamefont {L.}~\bibnamefont
  {Lazzarini}}, \bibinfo {author} {\bibfnamefont {R.}~\bibnamefont {Mantovan}},
  \bibinfo {author} {\bibfnamefont {F.}~\bibnamefont {Maccherozzi}}, \bibinfo
  {author} {\bibfnamefont {S.~S.}\ \bibnamefont {Dhesi}}, \bibinfo {author}
  {\bibfnamefont {B.}~\bibnamefont {Ocker}}, \bibinfo {author} {\bibfnamefont
  {C.~H.}\ \bibnamefont {Marrows}}, \bibinfo {author} {\bibfnamefont {T.~A.}\
  \bibnamefont {Moore}}, \ and\ \bibinfo {author} {\bibfnamefont
  {M.}~\bibnamefont {Kl\"aui}},\ }\href {\doibase 10.1103/PhysRevB.91.014433}
  {\bibfield  {journal} {\bibinfo  {journal} {Phys. Rev. B}\ }\textbf {\bibinfo
  {volume} {91}},\ \bibinfo {pages} {014433} (\bibinfo {year}
  {2015})}\BibitemShut {NoStop}%
\bibitem [{\citenamefont {Hrabec}\ \emph {et~al.}(2014)\citenamefont {Hrabec},
  \citenamefont {Porter}, \citenamefont {Wells}, \citenamefont {Benitez},
  \citenamefont {Burnell}, \citenamefont {McVitie}, \citenamefont {McGrouther},
  \citenamefont {Moore},\ and\ \citenamefont {Marrows}}]{hrabec2014measuring}%
  \BibitemOpen
  \bibfield  {author} {\bibinfo {author} {\bibfnamefont {A.}~\bibnamefont
  {Hrabec}}, \bibinfo {author} {\bibfnamefont {N.}~\bibnamefont {Porter}},
  \bibinfo {author} {\bibfnamefont {A.}~\bibnamefont {Wells}}, \bibinfo
  {author} {\bibfnamefont {M.}~\bibnamefont {Benitez}}, \bibinfo {author}
  {\bibfnamefont {G.}~\bibnamefont {Burnell}}, \bibinfo {author} {\bibfnamefont
  {S.}~\bibnamefont {McVitie}}, \bibinfo {author} {\bibfnamefont
  {D.}~\bibnamefont {McGrouther}}, \bibinfo {author} {\bibfnamefont
  {T.}~\bibnamefont {Moore}}, \ and\ \bibinfo {author} {\bibfnamefont
  {C.}~\bibnamefont {Marrows}},\ }\href@noop {} {\bibfield  {journal} {\bibinfo
   {journal} {Phys. Rev. B}\ }\textbf {\bibinfo {volume} {90}},\ \bibinfo
  {pages} {020402} (\bibinfo {year} {2014})}\BibitemShut {NoStop}%
\bibitem [{\citenamefont {Emori}\ \emph {et~al.}(2013)\citenamefont {Emori},
  \citenamefont {Bauer}, \citenamefont {Ahn}, \citenamefont {Martinez},\ and\
  \citenamefont {Beach}}]{emori2013current}%
  \BibitemOpen
  \bibfield  {author} {\bibinfo {author} {\bibfnamefont {S.}~\bibnamefont
  {Emori}}, \bibinfo {author} {\bibfnamefont {U.}~\bibnamefont {Bauer}},
  \bibinfo {author} {\bibfnamefont {S.}~\bibnamefont {Ahn}}, \bibinfo {author}
  {\bibfnamefont {E.}~\bibnamefont {Martinez}}, \ and\ \bibinfo {author}
  {\bibfnamefont {G.}~\bibnamefont {Beach}},\ }\href@noop {} {\bibfield
  {journal} {\bibinfo  {journal} {Nat. Mater.}\ }\textbf {\bibinfo {volume}
  {12}},\ \bibinfo {pages} {611} (\bibinfo {year} {2013})}\BibitemShut
  {NoStop}%
\bibitem [{\citenamefont {Ryu}\ \emph {et~al.}(2013)\citenamefont {Ryu},
  \citenamefont {Thomas}, \citenamefont {Yang},\ and\ \citenamefont
  {Parkin}}]{ryu2013chiral}%
  \BibitemOpen
  \bibfield  {author} {\bibinfo {author} {\bibfnamefont {K.-S.}\ \bibnamefont
  {Ryu}}, \bibinfo {author} {\bibfnamefont {L.}~\bibnamefont {Thomas}},
  \bibinfo {author} {\bibfnamefont {S.-H.}\ \bibnamefont {Yang}}, \ and\
  \bibinfo {author} {\bibfnamefont {S.}~\bibnamefont {Parkin}},\ }\href@noop {}
  {\bibfield  {journal} {\bibinfo  {journal} {Nat. Nanotechnol.}\ }\textbf
  {\bibinfo {volume} {8}},\ \bibinfo {pages} {527} (\bibinfo {year}
  {2013})}\BibitemShut {NoStop}%
\bibitem [{\citenamefont {Woo}\ \emph {et~al.}(2016)\citenamefont {Woo},
  \citenamefont {Litzius}, \citenamefont {Kr{\"u}ger}, \citenamefont {Im},
  \citenamefont {Caretta}, \citenamefont {Richter}, \citenamefont {Mann},
  \citenamefont {Krone}, \citenamefont {Reeve}, \citenamefont {Weigand},
  \citenamefont {Agrawal}, \citenamefont {Lemesh}, \citenamefont {Mawass},
  \citenamefont {Fischer}, \citenamefont {Kläui},\ and\ \citenamefont
  {Beach}}]{woo2016}%
  \BibitemOpen
  \bibfield  {author} {\bibinfo {author} {\bibfnamefont {S.}~\bibnamefont
  {Woo}}, \bibinfo {author} {\bibfnamefont {K.}~\bibnamefont {Litzius}},
  \bibinfo {author} {\bibfnamefont {B.}~\bibnamefont {Kr{\"u}ger}}, \bibinfo
  {author} {\bibfnamefont {M.-Y.}\ \bibnamefont {Im}}, \bibinfo {author}
  {\bibfnamefont {L.}~\bibnamefont {Caretta}}, \bibinfo {author} {\bibfnamefont
  {K.}~\bibnamefont {Richter}}, \bibinfo {author} {\bibfnamefont
  {M.}~\bibnamefont {Mann}}, \bibinfo {author} {\bibfnamefont {A.}~\bibnamefont
  {Krone}}, \bibinfo {author} {\bibfnamefont {R.~M.}\ \bibnamefont {Reeve}},
  \bibinfo {author} {\bibfnamefont {M.}~\bibnamefont {Weigand}}, \bibinfo
  {author} {\bibfnamefont {P.}~\bibnamefont {Agrawal}}, \bibinfo {author}
  {\bibfnamefont {I.}~\bibnamefont {Lemesh}}, \bibinfo {author} {\bibfnamefont
  {M.}~\bibnamefont {Mawass}}, \bibinfo {author} {\bibfnamefont
  {P.}~\bibnamefont {Fischer}}, \bibinfo {author} {\bibfnamefont
  {M.}~\bibnamefont {Kläui}}, \ and\ \bibinfo {author} {\bibfnamefont
  {G.}~\bibnamefont {Beach}},\ }\href@noop {} {\bibfield  {journal} {\bibinfo
  {journal} {Nat. Mater.}\ }\textbf {\bibinfo {volume} {15}},\ \bibinfo {pages}
  {501} (\bibinfo {year} {2016})}\BibitemShut {NoStop}%
\bibitem [{\citenamefont {Jiang}\ \emph {et~al.}(2015)\citenamefont {Jiang},
  \citenamefont {Upadhyaya}, \citenamefont {Zhang}, \citenamefont {Yu},
  \citenamefont {Jungfleisch}, \citenamefont {Fradin}, \citenamefont {Pearson},
  \citenamefont {Tserkovnyak}, \citenamefont {Wang}, \citenamefont {Heinonen},
  \citenamefont {te~Velthuis},\ and\ \citenamefont {Hoffmann}}]{Jiang283}%
  \BibitemOpen
  \bibfield  {author} {\bibinfo {author} {\bibfnamefont {W.}~\bibnamefont
  {Jiang}}, \bibinfo {author} {\bibfnamefont {P.}~\bibnamefont {Upadhyaya}},
  \bibinfo {author} {\bibfnamefont {W.}~\bibnamefont {Zhang}}, \bibinfo
  {author} {\bibfnamefont {G.}~\bibnamefont {Yu}}, \bibinfo {author}
  {\bibfnamefont {M.~B.}\ \bibnamefont {Jungfleisch}}, \bibinfo {author}
  {\bibfnamefont {F.~Y.}\ \bibnamefont {Fradin}}, \bibinfo {author}
  {\bibfnamefont {J.~E.}\ \bibnamefont {Pearson}}, \bibinfo {author}
  {\bibfnamefont {Y.}~\bibnamefont {Tserkovnyak}}, \bibinfo {author}
  {\bibfnamefont {K.~L.}\ \bibnamefont {Wang}}, \bibinfo {author}
  {\bibfnamefont {O.}~\bibnamefont {Heinonen}}, \bibinfo {author}
  {\bibfnamefont {S.~G.~E.}\ \bibnamefont {te~Velthuis}}, \ and\ \bibinfo
  {author} {\bibfnamefont {A.}~\bibnamefont {Hoffmann}},\ }\href@noop {}
  {\bibfield  {journal} {\bibinfo  {journal} {Science}\ }\textbf {\bibinfo
  {volume} {349}},\ \bibinfo {pages} {283} (\bibinfo {year}
  {2015})}\BibitemShut {NoStop}%
\bibitem [{\citenamefont {Zhang}\ \emph {et~al.}(2015)\citenamefont {Zhang},
  \citenamefont {Zhao}, \citenamefont {Fangohr}, \citenamefont {Liu},
  \citenamefont {Xia}, \citenamefont {Xia},\ and\ \citenamefont
  {Morvan}}]{zhang_skyrmion-skyrmion_2015}%
  \BibitemOpen
  \bibfield  {author} {\bibinfo {author} {\bibfnamefont {X.}~\bibnamefont
  {Zhang}}, \bibinfo {author} {\bibfnamefont {G.~P.}\ \bibnamefont {Zhao}},
  \bibinfo {author} {\bibfnamefont {H.}~\bibnamefont {Fangohr}}, \bibinfo
  {author} {\bibfnamefont {J.~P.}\ \bibnamefont {Liu}}, \bibinfo {author}
  {\bibfnamefont {W.~X.}\ \bibnamefont {Xia}}, \bibinfo {author} {\bibfnamefont
  {J.}~\bibnamefont {Xia}}, \ and\ \bibinfo {author} {\bibfnamefont {F.~J.}\
  \bibnamefont {Morvan}},\ }\href {\doibase 10.1038/srep07643} {\bibfield
  {journal} {\bibinfo  {journal} {Sci. Rep.}\ }\textbf {\bibinfo {volume}
  {5}},\ \bibinfo {pages} {7643} (\bibinfo {year} {2015})}\BibitemShut
  {NoStop}%
\bibitem [{\citenamefont {Everschor-Sitte}\ and\ \citenamefont
  {Sitte}(2014)}]{everschor2014real}%
  \BibitemOpen
  \bibfield  {author} {\bibinfo {author} {\bibfnamefont {K.}~\bibnamefont
  {Everschor-Sitte}}\ and\ \bibinfo {author} {\bibfnamefont {M.}~\bibnamefont
  {Sitte}},\ }\href@noop {} {\bibfield  {journal} {\bibinfo  {journal} {J.
  Appl. Phys.}\ }\textbf {\bibinfo {volume} {115}},\ \bibinfo {pages} {172602}
  (\bibinfo {year} {2014})}\BibitemShut {NoStop}%
\bibitem [{\citenamefont {Martinez}\ and\ \citenamefont
  {Alejos}(2014)}]{martinez_coupled_2014}%
  \BibitemOpen
  \bibfield  {author} {\bibinfo {author} {\bibfnamefont {E.}~\bibnamefont
  {Martinez}}\ and\ \bibinfo {author} {\bibfnamefont {O.}~\bibnamefont
  {Alejos}},\ }\href {\doibase 10.1063/1.4889848} {\bibfield  {journal}
  {\bibinfo  {journal} {J. Appl. Phys.}\ }\textbf {\bibinfo {volume} {116}},\
  \bibinfo {pages} {023909} (\bibinfo {year} {2014})}\BibitemShut {NoStop}%
\bibitem [{\citenamefont {Hiramatsu}\ \emph {et~al.}(2014)\citenamefont
  {Hiramatsu}, \citenamefont {Kim}, \citenamefont {Nakatani}, \citenamefont
  {Moriyama},\ and\ \citenamefont {Ono}}]{hiramatsu2014}%
  \BibitemOpen
  \bibfield  {author} {\bibinfo {author} {\bibfnamefont {R.}~\bibnamefont
  {Hiramatsu}}, \bibinfo {author} {\bibfnamefont {K.-J.}\ \bibnamefont {Kim}},
  \bibinfo {author} {\bibfnamefont {Y.}~\bibnamefont {Nakatani}}, \bibinfo
  {author} {\bibfnamefont {T.}~\bibnamefont {Moriyama}}, \ and\ \bibinfo
  {author} {\bibfnamefont {T.}~\bibnamefont {Ono}},\ }\href@noop {} {\bibfield
  {journal} {\bibinfo  {journal} {Jpn. J. Appl. Phys.}\ }\textbf {\bibinfo
  {volume} {53}},\ \bibinfo {pages} {108001} (\bibinfo {year}
  {2014})}\BibitemShut {NoStop}%
\bibitem [{\citenamefont {Benitez}\ \emph {et~al.}(2015)\citenamefont
  {Benitez}, \citenamefont {Hrabec}, \citenamefont {Mihai}, \citenamefont
  {Moore}, \citenamefont {Burnell}, \citenamefont {McGrouther}, \citenamefont
  {Marrows},\ and\ \citenamefont {McVitie}}]{benitez2015}%
  \BibitemOpen
  \bibfield  {author} {\bibinfo {author} {\bibfnamefont {M.}~\bibnamefont
  {Benitez}}, \bibinfo {author} {\bibfnamefont {A.}~\bibnamefont {Hrabec}},
  \bibinfo {author} {\bibfnamefont {A.}~\bibnamefont {Mihai}}, \bibinfo
  {author} {\bibfnamefont {T.}~\bibnamefont {Moore}}, \bibinfo {author}
  {\bibfnamefont {G.}~\bibnamefont {Burnell}}, \bibinfo {author} {\bibfnamefont
  {D.}~\bibnamefont {McGrouther}}, \bibinfo {author} {\bibfnamefont
  {C.}~\bibnamefont {Marrows}}, \ and\ \bibinfo {author} {\bibfnamefont
  {S.}~\bibnamefont {McVitie}},\ }\href@noop {} {\bibfield  {journal} {\bibinfo
   {journal} {Nat. Commun.}\ }\textbf {\bibinfo {volume} {6}},\ \bibinfo
  {pages} {8957} (\bibinfo {year} {2015})}\BibitemShut {NoStop}%
\bibitem [{\citenamefont {P.~del Real}\ \emph {et~al.}(2017)\citenamefont
  {P.~del Real}, \citenamefont {Raposo}, \citenamefont {Martinez},\ and\
  \citenamefont {Hayashi}}]{del_real_current-induced_2017}%
  \BibitemOpen
  \bibfield  {author} {\bibinfo {author} {\bibfnamefont {R.}~\bibnamefont
  {P.~del Real}}, \bibinfo {author} {\bibfnamefont {V.}~\bibnamefont {Raposo}},
  \bibinfo {author} {\bibfnamefont {E.}~\bibnamefont {Martinez}}, \ and\
  \bibinfo {author} {\bibfnamefont {M.}~\bibnamefont {Hayashi}},\ }\href@noop
  {} {\bibfield  {journal} {\bibinfo  {journal} {Nano Lett.}\ }\textbf
  {\bibinfo {volume} {17}},\ \bibinfo {pages} {1814} (\bibinfo {year}
  {2017})}\BibitemShut {NoStop}%
\bibitem [{\citenamefont {Torrejon}\ \emph {et~al.}(2014)\citenamefont
  {Torrejon}, \citenamefont {Kim}, \citenamefont {Sinha}, \citenamefont
  {Mitani}, \citenamefont {Hayashi}, \citenamefont {Yamanouchi},\ and\
  \citenamefont {Ohno}}]{torrejon_interface_2014}%
  \BibitemOpen
  \bibfield  {author} {\bibinfo {author} {\bibfnamefont {J.}~\bibnamefont
  {Torrejon}}, \bibinfo {author} {\bibfnamefont {J.}~\bibnamefont {Kim}},
  \bibinfo {author} {\bibfnamefont {J.}~\bibnamefont {Sinha}}, \bibinfo
  {author} {\bibfnamefont {S.}~\bibnamefont {Mitani}}, \bibinfo {author}
  {\bibfnamefont {M.}~\bibnamefont {Hayashi}}, \bibinfo {author} {\bibfnamefont
  {M.}~\bibnamefont {Yamanouchi}}, \ and\ \bibinfo {author} {\bibfnamefont
  {H.}~\bibnamefont {Ohno}},\ }\href {\doibase 10.1038/ncomms5655} {\bibfield
  {journal} {\bibinfo  {journal} {Nat. Commun.}\ }\textbf {\bibinfo {volume}
  {5}},\ \bibinfo {pages} {4655} (\bibinfo {year} {2014})}\BibitemShut
  {NoStop}%
\bibitem [{\citenamefont {Je}\ \emph {et~al.}(2013)\citenamefont {Je},
  \citenamefont {Kim}, \citenamefont {Yoo}, \citenamefont {Min}, \citenamefont
  {Lee},\ and\ \citenamefont {Choe}}]{je_asymmetric_2013}%
  \BibitemOpen
  \bibfield  {author} {\bibinfo {author} {\bibfnamefont {S.-G.}\ \bibnamefont
  {Je}}, \bibinfo {author} {\bibfnamefont {D.-H.}\ \bibnamefont {Kim}},
  \bibinfo {author} {\bibfnamefont {S.-C.}\ \bibnamefont {Yoo}}, \bibinfo
  {author} {\bibfnamefont {B.-C.}\ \bibnamefont {Min}}, \bibinfo {author}
  {\bibfnamefont {K.-J.}\ \bibnamefont {Lee}}, \ and\ \bibinfo {author}
  {\bibfnamefont {S.-B.}\ \bibnamefont {Choe}},\ }\href {\doibase
  10.1103/PhysRevB.88.214401} {\bibfield  {journal} {\bibinfo  {journal} {Phys.
  Rev. B}\ }\textbf {\bibinfo {volume} {88}},\ \bibinfo {pages} {214401}
  (\bibinfo {year} {2013})}\BibitemShut {NoStop}%
\bibitem [{\citenamefont {Tarasenko}\ \emph {et~al.}(1998)\citenamefont
  {Tarasenko}, \citenamefont {Stankiewicz}, \citenamefont {Tarasenko},\ and\
  \citenamefont {Ferr{\'e}}}]{Tarasenko199819}%
  \BibitemOpen
  \bibfield  {author} {\bibinfo {author} {\bibfnamefont {S.}~\bibnamefont
  {Tarasenko}}, \bibinfo {author} {\bibfnamefont {A.}~\bibnamefont
  {Stankiewicz}}, \bibinfo {author} {\bibfnamefont {V.}~\bibnamefont
  {Tarasenko}}, \ and\ \bibinfo {author} {\bibfnamefont {J.}~\bibnamefont
  {Ferr{\'e}}},\ }\href@noop {} {\bibfield  {journal} {\bibinfo  {journal} {J.
  Magn. Magn. Mater.}\ }\textbf {\bibinfo {volume} {189}},\ \bibinfo {pages}
  {19} (\bibinfo {year} {1998})}\BibitemShut {NoStop}%
\bibitem [{\citenamefont {Leonov}\ and\ \citenamefont
  {Mostovoy}(2015)}]{leonov_multiply_2015}%
  \BibitemOpen
  \bibfield  {author} {\bibinfo {author} {\bibfnamefont {A.~O.}\ \bibnamefont
  {Leonov}}\ and\ \bibinfo {author} {\bibfnamefont {M.}~\bibnamefont
  {Mostovoy}},\ }\href@noop {} {\bibfield  {journal} {\bibinfo  {journal} {Nat.
  Commun.}\ }\textbf {\bibinfo {volume} {6}},\ \bibinfo {pages} {8275}
  (\bibinfo {year} {2015})}\BibitemShut {NoStop}%
\bibitem [{\citenamefont {Hurd}(1972)}]{hurd1972hall}%
  \BibitemOpen
  \bibfield  {author} {\bibinfo {author} {\bibfnamefont {C.}~\bibnamefont
  {Hurd}},\ }\href {https://books.google.de/books?id=DjFRAAAAMAAJ} {\emph
  {\bibinfo {title} {The Hall Effect in Metals and Alloys}}},\ International
  Congresses of Quantum Chemistry Series\ (\bibinfo  {publisher} {Plenum
  Press},\ \bibinfo {year} {1972})\BibitemShut {NoStop}%
\bibitem [{\citenamefont {Liu}\ \emph {et~al.}(2012)\citenamefont {Liu},
  \citenamefont {Pai}, \citenamefont {Li}, \citenamefont {Tseng}, \citenamefont
  {Ralph},\ and\ \citenamefont {Buhrman}}]{Liu555}%
  \BibitemOpen
  \bibfield  {author} {\bibinfo {author} {\bibfnamefont {L.}~\bibnamefont
  {Liu}}, \bibinfo {author} {\bibfnamefont {C.-F.}\ \bibnamefont {Pai}},
  \bibinfo {author} {\bibfnamefont {Y.}~\bibnamefont {Li}}, \bibinfo {author}
  {\bibfnamefont {H.~W.}\ \bibnamefont {Tseng}}, \bibinfo {author}
  {\bibfnamefont {D.~C.}\ \bibnamefont {Ralph}}, \ and\ \bibinfo {author}
  {\bibfnamefont {R.~A.}\ \bibnamefont {Buhrman}},\ }\href {\doibase
  10.1126/science.1218197} {\ \textbf {\bibinfo {volume} {336}},\ \bibinfo
  {pages} {555} (\bibinfo {year} {2012})}\BibitemShut {NoStop}%
\bibitem [{\citenamefont {Garello}\ \emph {et~al.}(2013)\citenamefont
  {Garello}, \citenamefont {Miron}, \citenamefont {Avci}, \citenamefont
  {Freimuth}, \citenamefont {Mokrousov}, \citenamefont {Bl{\"u}gel},
  \citenamefont {Auffret}, \citenamefont {Boulle}, \citenamefont {Gaudin},\
  and\ \citenamefont {Gambardella}}]{garello2013symmetry}%
  \BibitemOpen
  \bibfield  {author} {\bibinfo {author} {\bibfnamefont {K.}~\bibnamefont
  {Garello}}, \bibinfo {author} {\bibfnamefont {I.~M.}\ \bibnamefont {Miron}},
  \bibinfo {author} {\bibfnamefont {C.~O.}\ \bibnamefont {Avci}}, \bibinfo
  {author} {\bibfnamefont {F.}~\bibnamefont {Freimuth}}, \bibinfo {author}
  {\bibfnamefont {Y.}~\bibnamefont {Mokrousov}}, \bibinfo {author}
  {\bibfnamefont {S.}~\bibnamefont {Bl{\"u}gel}}, \bibinfo {author}
  {\bibfnamefont {S.}~\bibnamefont {Auffret}}, \bibinfo {author} {\bibfnamefont
  {O.}~\bibnamefont {Boulle}}, \bibinfo {author} {\bibfnamefont
  {G.}~\bibnamefont {Gaudin}}, \ and\ \bibinfo {author} {\bibfnamefont
  {P.}~\bibnamefont {Gambardella}},\ }\href@noop {} {\bibfield  {journal}
  {\bibinfo  {journal} {Nat. Nanotechnol.}\ }\textbf {\bibinfo {volume} {8}},\
  \bibinfo {pages} {587} (\bibinfo {year} {2013})}\BibitemShut {NoStop}%
\bibitem [{\citenamefont {Conte}\ \emph {et~al.}(2017)\citenamefont {Conte},
  \citenamefont {Karnad}, \citenamefont {Martinez}, \citenamefont {Lee},
  \citenamefont {Kim}, \citenamefont {Han}, \citenamefont {Kim}, \citenamefont
  {Prenzel}, \citenamefont {Schulz}, \citenamefont {You}, \citenamefont
  {Swagten},\ and\ \citenamefont {Kl\"aui}}]{LoConte_2017}%
  \BibitemOpen
  \bibfield  {author} {\bibinfo {author} {\bibfnamefont {R.~L.}\ \bibnamefont
  {Conte}}, \bibinfo {author} {\bibfnamefont {G.~V.}\ \bibnamefont {Karnad}},
  \bibinfo {author} {\bibfnamefont {E.}~\bibnamefont {Martinez}}, \bibinfo
  {author} {\bibfnamefont {K.}~\bibnamefont {Lee}}, \bibinfo {author}
  {\bibfnamefont {N.-H.}\ \bibnamefont {Kim}}, \bibinfo {author} {\bibfnamefont
  {D.-S.}\ \bibnamefont {Han}}, \bibinfo {author} {\bibfnamefont {J.-S.}\
  \bibnamefont {Kim}}, \bibinfo {author} {\bibfnamefont {S.}~\bibnamefont
  {Prenzel}}, \bibinfo {author} {\bibfnamefont {T.}~\bibnamefont {Schulz}},
  \bibinfo {author} {\bibfnamefont {C.-Y.}\ \bibnamefont {You}}, \bibinfo
  {author} {\bibfnamefont {H.~J.~M.}\ \bibnamefont {Swagten}}, \ and\ \bibinfo
  {author} {\bibfnamefont {M.}~\bibnamefont {Kl\"aui}},\ }\href {\doibase
  10.1063/1.4990694} {\bibfield  {journal} {\bibinfo  {journal} {AIP Adv.}\
  }\textbf {\bibinfo {volume} {7}},\ \bibinfo {pages} {065317} (\bibinfo {year}
  {2017})}\BibitemShut {NoStop}%
\bibitem [{\citenamefont {Kim}\ \emph {et~al.}(2011)\citenamefont {Kim},
  \citenamefont {Moon}, \citenamefont {Lee},\ and\ \citenamefont
  {Choe}}]{kim_control_2011}%
  \BibitemOpen
  \bibfield  {author} {\bibinfo {author} {\bibfnamefont {K.-J.}\ \bibnamefont
  {Kim}}, \bibinfo {author} {\bibfnamefont {K.-W.}\ \bibnamefont {Moon}},
  \bibinfo {author} {\bibfnamefont {K.-S.}\ \bibnamefont {Lee}}, \ and\
  \bibinfo {author} {\bibfnamefont {S.-B.}\ \bibnamefont {Choe}},\ }\href@noop
  {} {\bibfield  {journal} {\bibinfo  {journal} {Nanotechnology}\ }\textbf
  {\bibinfo {volume} {22}},\ \bibinfo {pages} {025702} (\bibinfo {year}
  {2011})}\BibitemShut {NoStop}%
\bibitem [{\citenamefont {Vansteenkiste}\ \emph {et~al.}(2014)\citenamefont
  {Vansteenkiste}, \citenamefont {Leliaert}, \citenamefont {Dvornik},
  \citenamefont {Helsen}, \citenamefont {Garcia-Sanchez},\ and\ \citenamefont
  {Waeyenberge}}]{Vansteenkiste_MuMax_2014}%
  \BibitemOpen
  \bibfield  {author} {\bibinfo {author} {\bibfnamefont {A.}~\bibnamefont
  {Vansteenkiste}}, \bibinfo {author} {\bibfnamefont {J.}~\bibnamefont
  {Leliaert}}, \bibinfo {author} {\bibfnamefont {M.}~\bibnamefont {Dvornik}},
  \bibinfo {author} {\bibfnamefont {M.}~\bibnamefont {Helsen}}, \bibinfo
  {author} {\bibfnamefont {F.}~\bibnamefont {Garcia-Sanchez}}, \ and\ \bibinfo
  {author} {\bibfnamefont {B.~V.}\ \bibnamefont {Waeyenberge}},\ }\href@noop {}
  {\bibfield  {journal} {\bibinfo  {journal} {AIP Adv.}\ }\textbf {\bibinfo
  {volume} {4}},\ \bibinfo {pages} {107133} (\bibinfo {year}
  {2014})}\BibitemShut {NoStop}%
\end{thebibliography}

%

\end{document}